\documentclass[10pt,conference]{IEEEtran}

\IEEEoverridecommandlockouts

\usepackage{listings}
\usepackage{tabularray}

\usepackage{cite}
\usepackage{amsmath,amssymb,amsfonts}
\usepackage{amsthm}
\usepackage{algorithmic}
\usepackage{graphicx}
\usepackage{textcomp}
\usepackage{xcolor}
\usepackage{url}
\def\BibTeX{{\rm B\kern-.05em{\sc i\kern-.025em b}\kern-.08em
    T\kern-.1667em\lower.7ex\hbox{E}\kern-.125emX}}

\theoremstyle{definition}
\newtheorem{definition}{Definition}

\usepackage{pbalance}

\makeatletter
\newcommand{\linebreakand}{%
  \end{@IEEEauthorhalign}
  \hfill\mbox{}\par
  \mbox{}\hfill\begin{@IEEEauthorhalign}
}
\makeatother

\newcommand{\ldpkup}{LDP$^3$~}

\begin{document}
\title{LDP$^3$: An Extensible and Multi-Threaded Toolkit for Local Differential Privacy Protocols and Post-Processing Methods}

\author{%
\IEEEauthorblockN{Berkay Kemal Balioglu}
\IEEEauthorblockA{\textit{Department of Computer Engineering} \\
\textit{Koç University}\\
Istanbul, Turkiye \\
bbalioglu23@ku.edu.tr}
\and
\IEEEauthorblockN{Alireza Khodaie}
\IEEEauthorblockA{\textit{Department of Computer Engineering} \\
\textit{Koç University}\\
Istanbul, Turkiye \\
akhodaie22@ku.edu.tr}
\and
\IEEEauthorblockN{M. Emre Gursoy}
\IEEEauthorblockA{\textit{Department of Computer Engineering} \\
\textit{Koç University}\\
Istanbul, Turkiye \\
emregursoy@ku.edu.tr}
}

\maketitle

\begin{abstract}
Local differential privacy (LDP) has become a prominent notion for privacy-preserving data collection. While numerous LDP protocols and post-processing (PP) methods have been developed, selecting an optimal combination under different privacy budgets and datasets remains a challenge. Moreover, the lack of a comprehensive and extensible LDP benchmarking toolkit raises difficulties in evaluating new protocols and PP methods. To address these concerns, this paper presents LDP$^3$ (pronounced LDP-Cube), an open-source, extensible, and multi-threaded toolkit for LDP researchers and practitioners. \ldpkup contains implementations of several LDP protocols, PP methods, and utility metrics in a modular and extensible design. Its modular design enables developers to conveniently integrate new protocols and PP methods. Furthermore, its multi-threaded nature enables significant reductions in execution times via parallelization. Experimental evaluations demonstrate that: (i) using \ldpkup to select a good protocol and post-processing method substantially improves utility compared to a bad or random choice, and (ii) the multi-threaded design of \ldpkup brings substantial benefits in terms of efficiency.
\end{abstract}


\section{Introduction} \label{sec:Introduction}



In recent years, local differential privacy (LDP) has emerged as a popular notion for privacy-preserving data collection. In LDP, each user perturbs their sensitive data on their device before sharing it with a central server. Since the privacy protection step occurs on the user's device through a randomized algorithm, LDP allows for privacy-preserving data analysis when users do not trust the central server (i.e., data collector). Considering that LDP offers privacy to users while enabling the data collector to estimate aggregate statistics, it has become popular in both research and industry \cite{cormode2018privacy,yang2023local}. For example, Google developed RAPPOR to analyze users' default browser homepages and search engines, Apple used LDP to identify popular emojis and trending words for typing recommendations, and Microsoft implemented LDP in Windows 10 to collect app usage telemetry \cite{thakurta2017emoji,ding2017collecting,erlingsson2014rappor}.

The popularity of LDP has led to the development of various LDP protocols, such as GRR, BLH, OLH, RAPPOR, OUE, and SS \cite{wang2017locally,gursoy2022adversarial,cormode2021frequency}. In addition, to improve the utility of server-side estimations, several post-processing (PP) methods have been proposed, such as Base-Pos, Norm, Norm-Cut, Norm-Sub, Norm-Mul, Power, and PowerNS \cite{jia2019calibrate,wang2020locally}. Yet, no standardized criteria exist for selecting an optimal combination of LDP protocol and PP method across all possible $\varepsilon$ and datasets. Furthermore, to advance development in LDP, it would be beneficial to have a toolkit where researchers and practitioners can conveniently integrate their new protocols and PP methods, and benchmark them against existing ones. The existence of such a toolkit would also provide guidance for practitioners in terms of which protocols and methods would yield the highest utility in their specific deployments.  


To address these needs, in this paper, we propose and develop \ldpkup (pronounced LDP-Cube, stands for: \textbf{L}ocal \textbf{D}ifferential \textbf{P}rivacy with \textbf{P}ost \textbf{P}rocessing). \ldpkup is an open-source toolkit\footnote{https://github.com/alrzakh/LDPcube} containing implementations of 6 LDP protocols, 7 post-processing methods, and several utility metrics in a multi-threaded setup. By housing many of the popular protocols and PP methods, \ldpkup aims to offer a holistic resource for researchers and practitioners to integrate and test their newly proposed methods, or to benchmark existing methods on new applications and datasets. The modular design of \ldpkup enables it to be an extensible privacy toolkit, allowing the addition of new protocols, methods, and metrics as needed. Furthermore, the multi-threaded nature of \ldpkup is particularly beneficial, since the execution and benchmarking of multiple protocols and PP methods with many repetitions (to combat LDP's randomness and achieve statistical significance) causes high execution times. The parallelization in \ldpkup helps to reduce execution times substantially. 



There are several potential uses of \ldpkup which makes it beneficial to advance LDP research (from a researcher's perspective) and deployment (from a practitioner's perspective). For example:
\begin{itemize}
    \item Say that a privacy researcher proposes a new LDP protocol or PP method. The researcher can implement their protocol or method into \ldpkup and experimentally compare it with existing methods. Furthermore, for a newly developed protocol, the best-performing PP method can be found by experimenting with LDP$^3$.
    \item Say that a practitioner wants to apply LDP to a real-world data collection task. The practitioner has a surrogate data sample, the $\varepsilon$ budget, and the utility metric in hand. Using LDP$^3$, the practitioner can find which LDP protocol and PP method yields the lowest utility loss and apply this combination to the real-world data collection task.
    \item Existing PP methods have so far only been validated using a limited number of LDP protocols and settings (e.g., only OLH in \cite{wang2020locally}). A broader and more holistic benchmarking of different protocol and PP method combinations can be performed using \ldpkup to validate (or challenge) previous findings.
\end{itemize}

The remainder of this paper is organized as follows. In Section \ref{sec:Background}, we provide the necessary LDP background and explain the differences between \ldpkup and other LDP-related libraries and toolkits. In Section \ref{sec:LDPCube}, we describe the design and current implementation of LDP$^3$, as well as how to use \ldpkup in practice. In Section \ref{sec:Experiments}, we experimentally show the benefits and contributions of \ldpkup from two perspectives: (i) using \ldpkup to select a good protocol and PP method combination indeed helps improve utility substantially, and (ii) multi-threaded design of \ldpkup brings substantial benefits in reducing execution times. Section \ref{sec:Conclusion} concludes the paper. 

\section{Background and Related Work} \label{sec:Background}

\subsection{LDP Background and Notation}

In a typical LDP setting, there exist multiple clients (users) and a data collector (server). Each user perturbs their data locally on their device before sending it to the server. After perturbed values are collected, the server performs aggregation and estimation to recover population-level statistics. Since each user's data is perturbed using a randomized LDP algorithm, the server cannot infer exact information about any specific user.

Let $\mathcal{U}$ denote the user population and $\mathcal{D}$ denote the domain of users' values. We denote values in the domain by $v \in \mathcal{D}$. For user $u \in \mathcal{U}$, we denote this user's true value by $v_u$. We denote the true frequency of $v$ by $f(v)$, its estimated frequency after LDP by $\hat{f}(v)$, and its post-processed frequency by $\tilde{f}(v)$. A randomized mechanism $\psi$ is said to satisfy $\varepsilon$-LDP if the following holds.

\begin{definition}[$\varepsilon$-LDP] \label{def:LDP}
A randomized mechanism $\psi$ satisfies $\varepsilon$-LDP, if and only if for any two values $v_1$, $v_2$ in $\mathcal{D}$:
\begin{equation} \label{eq:LDPeqn}
\forall y \in Range(\psi): ~~~~ \frac{\text{Pr}[\psi(v_1) = y]}{\text{Pr}[\psi(v_2) = y]} \leq e^{\varepsilon}
\end{equation}
where $Range(\psi)$ denotes the set of all possible outputs of $\psi$. 
\end{definition}

Here, $\varepsilon$ is the parameter that determines the strength of privacy protection. It is often called the \textit{privacy parameter} or \textit{privacy budget}. Smaller $\varepsilon$ yields stronger privacy. 

\subsection{Related Work}

The popularity of LDP has led to its application in a variety of contexts and data analysis tasks, such as high-dimensional data collection \cite{da2023felip,arcolezi2021random,zhang2018calm}, heavy hitter identification \cite{zhu2023heavy,wang2021locally}, set-valued data analysis \cite{wang2023locally,huang2024joint}, geospatial data analysis \cite{hong2022collecting,du2023ldptrace}, and deep learning \cite{wang2023ppefl,truex2020ldp}. LDP protocols and PP methods serve as the fundamental building blocks of many such applications. There have also been a few works which build practical systems or toolkits for the analysis of LDP protocols. However, \ldpkup has some key differences compared to them, which are described below.

Earlier works such as \cite{wang2017locally} and \cite{wang2020locally} focus solely on protocols without post-processing, or evaluate post-processing only on a single protocol. The \textsc{Pure-LDP} package was proposed in \cite{cormode2021frequency}; however, it contains a subset of the post-processing methods in \ldpkup and it does not support multi-threading. The \textsc{Multi-Freq-LDP} package was proposed in \cite{Arcolezi2022}; however, the main focus of \textsc{Multi-Freq-LDP} is multi-dimensional and longitudinal frequency estimation \cite{arcolezi2021random,arcolezi2022improving}. Hence, it does not contain post-processing methods. \textsc{LDPLens} was proposed in \cite{gursoy2022adversarial}; however, it is focused on protocols' adversarial analysis and does not contain post-processing. Overall, \ldpkup is novel in combining many LDP protocols and post-processing methods under a single umbrella in a multi-threaded setup. 



\section{\ldpkup Design and Implementation} \label{sec:LDPCube}


\ldpkup is an open-source toolkit designed to address the need for a systematic and comprehensive tool to evaluate combinations of different LDP protocols and PP methods. Built in a modular and extensible way, \ldpkup allows users to experimentally find the optimal LDP protocol and PP method for a given dataset, utility metric, and privacy budget $\varepsilon$. It consists of the following primary modules (i.e., components):
\begin{itemize}
    \item The Protocol Module contains implementations of multiple state-of-the-art LDP protocols such as GRR, BLH, OLH, RAPPOR, OUE, and SS. It offers a standardized interface, allowing developers to add new protocols to \ldpkup by implementing the protocols' user-side perturbation and server-side estimation functions.
    \item The Post-Processing Module contains implementations of PP methods which take as input the estimated frequencies $\hat{f}(v)$ and produce post-processed $\tilde{f}(v)$. In the current implementation of LDP$^3$, several PP methods are already included. The module is designed in an extensible way so that new PP methods can be added in the future.
    \item The Utility Measurement Module contains implementations of utility metrics to measure the differences between $f(v)$ and $\hat{f}(v)$, or $f(v)$ and $\tilde{f}(v)$. Examples of currently implemented metrics include $\ell_1$ distance, $\ell_2$ distance, and KL-divergence. New metrics can be added as desired.
    \item The Multi-Threading Module aims to address high execution times when running many combinations of protocols and PP methods. It leverages Python's concurrency features to parallelize the data perturbation and aggregation processes, and significantly improves the computational efficiency of large-scale experiments. 
    \item Finally, the Execution Module provides a command-line interface to run experiments, saving detailed results to an output file or displaying aggregate results on the terminal. This module also facilitates the handling of datasets, e.g., reading and writing to files with standard formats (such as txt or csv).  
\end{itemize}
In the rest of this section, we describe the design and current implementation of each module in more detail. 

\subsection{Protocol Module} \label{sec:LDPprotocols}

Several LDP protocols have been developed in the literature \cite{cormode2021frequency,gursoy2022adversarial,wang2017locally}. These protocols are often used as building blocks in more complex data analysis tasks and downstream applications. An LDP protocol can be characterized by two main components: (i) user-side encoding and perturbation to satisfy LDP, and (ii) server-side aggregation and estimation to recover population-level statistics. \ldpkup currently contains the implementations of six protocols: GRR, RAPPOR, OUE, BLH, OLH, and SS. New protocols can be added to \ldpkup by implementing two functions: one function for the protocol's user-side encoding and perturbation, and one function for the protocol's server-side aggregation and estimation. The details of the protocols are provided below.

\textbf{Generalized Randomized Response (GRR)}. Randomized response is a method originally introduced for survey data collection, and GRR is an extension of this method designed for LDP. It allows for multi-valued domains (i.e., $|\mathcal{D}| \geq 3$) and works for any privacy parameter $\varepsilon$. Given a user's true value $v_u$, GRR generates a perturbed value $y_u \in \mathcal{D}$ based on the following probabilities: 
\begin{equation} \label{eq:perturbGRR}
    \Pr[\psi(v_u) = y_u] = 
    \begin{cases}
        p = \frac{e^\varepsilon}{e^\varepsilon + |\mathcal{D}| - 1}, & \text{if } y_u = v_u \\
        q = \frac{1}{e^\varepsilon + |\mathcal{D}| - 1}, & \text{if } y_u \neq v_u
    \end{cases}
\end{equation}

Once the server receives perturbed values from all users, it estimates the frequency of a specific value $v \in \mathcal{D}$ by first calculating $\widehat{C}(v)$, which is the number of users who reported $y_u = v$ as their perturbed value. Then, the estimate $\hat{f}(v)$ is found by:
\begin{equation} \label{eq:estimateGRR}
\hat{f}(v) = \frac{\widehat{C}(v) - |\mathcal{U}| \cdot q}{(p-q) \cdot |\mathcal{U}|}
\end{equation}

\textbf{RAPPOR.} The Randomized Aggregatable Privacy-Preserving Ordinal Response (RAPPOR) protocol, introduced by Google, ensures LDP by encoding a user's value into a bitvector and applying randomized perturbation. Although the original RAPPOR protocol uses Bloom filters for encoding, we describe a simpler version of RAPPOR with unary encoding, which is commonly used in the literature. Each user $u$ initializes a bitvector $B_u$ of length $|\mathcal{D}|$, setting all bits to 0 except for the one corresponding to the user's true value $v_u$: $B_u[v_u] = 1$. The RAPPOR perturbation mechanism then iterates through each bit $i \in [1, |B_u|]$ and keeps or flips the bit with probabilities defined in the following equation:
\begin{equation} \label{eq:RAPPORperturbation}
    \forall_{i \in [1,|B_u|]} : ~ \Pr[B_{u}'[i]=1] =
    \begin{cases}
        \frac{e^{\varepsilon/2}}{e^{\varepsilon/2} + 1}, & \text{if } B_{u}[i] = 1 \\
        \frac{1}{e^{\varepsilon/2} + 1}, & \text{if } B_{u}[i] = 0
    \end{cases}
\end{equation}
The perturbed bitvector $B_u'$ is sent to the server. After receiving perturbed bitvectors from all users, the server calculates $\widehat{C}[v]$, the count of 1’s at index $v$ across all received bitvectors:
\begin{equation} \label{eq:tildeC}
\widehat{C}[v] = \sum\limits_{u \in \mathcal{U}} B'_u[v]
\end{equation}
Finally, the server computes the estimate $\hat{f}(v)$ using the formula:
\begin{equation} \label{eq:estimateRAPPOR}
\hat{f}(v) = \frac{\widehat{C}[v] + |\mathcal{U}|\cdot(\alpha-1)}{(2\alpha-1) \cdot |\mathcal{U}|}
\end{equation}
where $\alpha$ is the bit-keeping probability:
$\alpha = \frac{e^{\varepsilon/2}}{e^{\varepsilon/2}+1}$.

\textbf{Optimized Unary Encoding (OUE)}. In OUE, after initializing the bitvector $B_u$ in the same way as in RAPPOR, the perturbed bitvector $B_u'$ is determined according to the following probabilities. These probabilities were mathematically derived to minimize the variance of server-side estimation \cite{wang2017locally}, improving protocol utility.
\begin{equation} \label{eq:OUEperturbation}
\forall_{i \in [1,|B_u|]} : ~ \Pr[B_u'[i]=1] =
\begin{cases}
\frac{1}{2}, & \text{if } B_u[i]=1 \\
\frac{1}{e^{\varepsilon}+1}, & \text{if } B_u[i]=0
\end{cases}
\end{equation}
The perturbed bitvector $B_u'$ is sent to the server. After receiving perturbed bitvectors from all users, the server calculates the count $\widehat{C}[v]$ in the same way as in Equation \ref{eq:tildeC}. The server then computes the estimate $\hat{f}(v)$ using the following formula:
\begin{equation}
\hat{f}(v) = \frac{2 \cdot \left( (e^\varepsilon+1) \cdot \widehat{C}[v] - |\mathcal{U}| \right)}{(e^\varepsilon - 1) \cdot |\mathcal{U}|}
\end{equation}

\textbf{Binary Local Hashing (BLH)}. Both RAPPOR and OUE utilize bitvectors of length $|\mathcal{D}|$, which can lead to significant user-side computation costs and user-server communication costs when $|\mathcal{D}|$ is large. BLH addresses these costs by applying hash functions to reduce the domain size. To address potential issues with hash collisions, BLH uses a set of hash functions $\mathcal{H}$, from which each user selects a different function.

Let $\mathcal{H}$ represent a set of hash functions such that each $H \in \mathcal{H}$ maps a value from $\mathcal{D}$ to an integer in the set $\{0, 1\}$, i.e., $H: \mathcal{D} \to \{0, 1\}$. Each user $u$ with true value $v_u$ randomly selects a hash function $H_u$ from $\mathcal{H}$ and computes the integer $x_u = H_u(v_u)$. The perturbation step in BLH then takes $x_u$ and produces a perturbed value $x'_u \in \{0, 1\}$ according to the following probabilities:
\begin{equation} \label{eq:BLHperturbation}
\forall_{i \in \{0, 1\}} : \text{Pr}[x'_u = i] =
\begin{cases}
\frac{e^{\varepsilon}}{e^{\varepsilon} + 1} & \text{if } x_u = i \\
\frac{1}{e^{\varepsilon} + 1} & \text{if } x_u \neq i
\end{cases}
\end{equation}
The user sends the tuple $\langle H_u, x'_u \rangle$ to the server. Once the server receives tuples from all users $u \in \mathcal{U}$, it estimates the value $v$ by first computing $Sup(v)$, which is the total count of tuples where the condition $x'_u = H_u(v)$ holds. The server then estimates $\hat{f}(v)$ as:
\begin{equation}
\hat{f}(v) = \frac{(e^{\varepsilon}+1) \cdot (2 \cdot Sup(v) - |\mathcal{U}|)}{(e^{\varepsilon}-1) \cdot |\mathcal{U}|}
\end{equation}

\textbf{Optimized Local Hashing (OLH)}. Unlike BLH, OLH uses a non-binary output space for hash functions. It allows the encoding of a value $v$ into an integer within the range $[0, g-1]$, where $g \geq 2$ is a parameter of the protocol. The motivation behind this modification is to overcome the utility loss that occurs in BLH when binary encoding is not optimal. OLH has been shown to provide significant utility improvements over BLH, particularly when $\varepsilon$ and $|\mathcal{U}|$ are large. The default value for $g$ is derived as $g = e^{\varepsilon} + 1$ \cite{wang2017locally}.

Let $\mathcal{H}$ represent a set of hash functions such that each $H \in \mathcal{H}$ maps a value from $\mathcal{D}$ to an integer in the range $[0, ~g-1]$, i.e., $H: \mathcal{D} \to [0, ~g-1]$. Each user randomly selects a hash function $H_u$ from $\mathcal{H}$ and computes the integer $x_u = H_u(v_u)$. The perturbation step in OLH then takes $x_u$ and produces a perturbed value $x'_u \in [0, g-1]$ with the following probabilities:
\begin{equation}
\forall i \in [0, g-1] : \text{Pr}[x'_u = i] =
\begin{cases}
\frac{e^{\varepsilon}}{e^{\varepsilon} + g - 1} & \text{if } x_u = i \\
\frac{1}{e^{\varepsilon} + g - 1} & \text{if } x_u \neq i
\end{cases}
\end{equation}
The user sends the tuple $\langle H_u, x'_u \rangle$ to the server. Once the server receives tuples from all users $u \in \mathcal{U}$, it estimates the value $v$ by first computing $Sup(v)$, which is the total count of tuples where the condition $x'_u = H_u(v)$ holds. The server then estimates $\hat{f}(v)$ as:
\begin{equation}
\hat{f}(v) = \frac{(e^{\varepsilon} + g - 1) \cdot (g \cdot Sup(v) - |\mathcal{U}|)}{(e^{\varepsilon} - 1) \cdot (g - 1) \cdot |\mathcal{U}|}
\end{equation}

\textbf{Subset Selection (SS)}. In the SS protocol, each user reports a subset $Z_u$ of the domain $\mathcal{D}$ to the server. The size of the subset $k = |Z_u|$ is a crucial parameter of the protocol. The default value for $k$ is defined as $k = \frac{|\mathcal{D}|}{e^{\varepsilon} + 1}$. User $u$ initializes subset $Z_u$ as empty. SS adds $v_u$ to $Z_u$ with probability $\frac{k \cdot e^{\varepsilon}}{k \cdot e^{\varepsilon} + |\mathcal{D}| - k}$. The remainder of the subset $Z_u$ is constructed as follows:
\begin{itemize}
    \item If $v_u$ was added to $Z_u$ in the previous step, then $k - 1$ elements are selected uniformly at random without replacement from $\mathcal{D} \setminus \{v_u\}$ and added to $Z_u$.
    \item If $v_u$ was not added to $Z_u$ in the previous step, then $k$ elements are selected uniformly at random without replacement from $\mathcal{D} \setminus \{v_u\}$ and added to $Z_u$.
\end{itemize}
The user sends the resulting $Z_u$ to the server. The server receives the subsets $Z_u$ from all users $u \in \mathcal{U}$. The server defines the parameters $\sigma_k$ and $\theta_k$ as:
\begin{equation}
\sigma_k = \frac{k \cdot e^{\varepsilon}}{k \cdot e^{\varepsilon} + |\mathcal{D}| - k}
\end{equation}
\begin{equation}
\theta_k = \frac{(k - 1) \cdot k \cdot e^{\varepsilon} + (|\mathcal{D}| - k) \cdot k}{(|\mathcal{D}| - 1) \cdot (k \cdot e^{\varepsilon} + |\mathcal{D}| - k)}
\end{equation}
To estimate $\hat{f}(v)$, the server computes $Sup(v)$, which is the total number of clients in $\mathcal{U}$ whose reported subset $Z_u$ contains $v$. Then, the server estimates $\hat{f}(v)$ as:
\begin{equation}
\hat{f}(v) = \frac{Sup(v) - |\mathcal{U}| \cdot \theta_k}{(\sigma_k - \theta_k) \cdot |\mathcal{U}|}
\end{equation}

\subsection{Post-Processing Module} \label{sec:postprocess}

Post-processing (PP) methods take as input the estimated frequencies under LDP $\hat{f}(v)$ and produce post-processed frequencies $\tilde{f}(v)$, with the aim of achieving consistency and utility improvement \cite{jia2019calibrate,wang2020locally}. Different PP methods result in varying trade-offs between increase in utility and bias; furthermore, different PP methods may be desirable for different datasets and LDP protocols. To facilitate the utilization and benchmarking of various PP methods, \ldpkup includes implementations of commonly used PP methods from the literature, which are described below. New methods can be added to the toolkit in the future. 

\textbf{Base-Pos:} Frequencies must be non-negative by definition; however, due to the randomization in LDP, $\hat{f}(v)$ may be negative for some $v \in \mathcal{D}$. Base-Pos addresses this problem by converting all negative estimations to 0. 
\begin{equation}
    \forall v \in \mathcal{D}:
\tilde{f}(v) =
\begin{cases} 
\hat{f}(v) & \text{if } \hat{f}(v) \geq 0 \\
0 & \text{otherwise}
\end{cases}
\end{equation}

\textbf{Norm:} The sum of frequencies across all $v \in \mathcal{D}$ should equal 1; however, this may not hold due to the randomization in LDP. Norm addresses this problem by adding a constant $\sigma$ to each frequency so that the sum will equal 1.
\begin{equation}
    \forall v \in \mathcal{D}:
\tilde{f}(v) = \hat{f}(v) + \sigma, \quad \text{such that} \quad \sum_{v \in \mathcal{D}} \tilde{f}(v) = 1
\end{equation}

\textbf{Norm-Cut:} Norm-Cut converts negative and small positive frequencies to 0, and it also ensures that the sum of frequencies equals 1. That is:
\begin{equation}
    \forall v \in \mathcal{D}:
\tilde{f}(v) =
\begin{cases}
0 & \text{if } \hat{f}(v) \leq \theta \\
\hat{f}(v) & \text{if } \hat{f}(v) > \theta
\end{cases}
\end{equation}
where $\theta$ is a threshold value. The value of $\theta$ is chosen such that $\sum_{v \in \mathcal{D}} \tilde{f}(v) = 1$ is ensured. 

\textbf{Norm-Sub:} Norm-Sub converts negative frequencies to 0. Then, it adds a constant $\delta$ to the frequencies to ensure that the sum of frequencies equals 1.
\begin{equation}
    \forall v \in \mathcal{D}:
\tilde{f}(v) =
\begin{cases}
0 & \text{if } \hat{f}(v) < 0 \\
\hat{f}(v) + \delta & \text{if } \hat{f}(v) \geq 0
\end{cases}
\end{equation}
Here, the value of $\delta$ is chosen such that $\sum_{v \in \mathcal{D}} \tilde{f}(v) = 1$ is ensured. 

\begin{figure*}[!t]
    \centering
    \includegraphics[width=.85\textwidth]{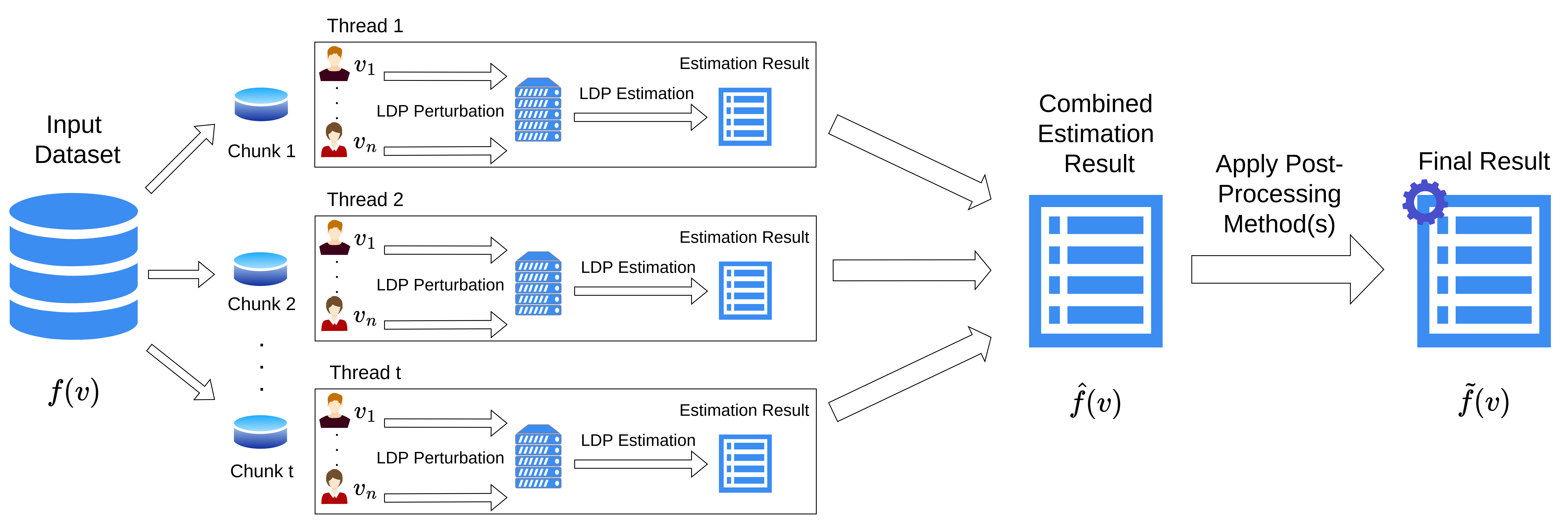}
    \caption{Multi-threaded execution in \ldpkup}
    \label{fig:multithread_system}
\end{figure*}

\textbf{Norm-Mul:} Norm-Mul converts negative frequencies to 0. Then, instead of an additive factor, it uses a multiplicative factor to the remaining frequencies so that their sum is 1.
\begin{equation}
    \forall v \in \mathcal{D}:
\tilde{f}(v) =
\begin{cases}
0 & \text{if } \hat{f}(v) < 0 \\
\alpha \hat{f}(v) & \text{if } \hat{f}(v) \geq 0
\end{cases}
\end{equation}
Here, $\alpha$ is the multiplicative factor. Its value is chosen such that $\sum_{v \in \mathcal{D}} \tilde{f}(v) = 1$ is ensured.

\textbf{Power:} The rationale of Power is that many real-world datasets follow a statistical distribution such as a Gaussian or power law distribution. Therefore, Power fits a distribution to the estimated frequencies and aims to minimize the expected square error. For example, with $\tilde{f}(v) \sim P$ where $P$ is a power law distribution, the goal is to minimize:
\begin{equation}
    \min_{P} ~ \mathbb{E}\left[ \sum_{v \in \mathcal{D}} \left( \hat{f}(v) - \tilde{f}(v) \right)^2 \right]
\end{equation}

\textbf{PowerNS:} PowerNS first applies Power and then uses Norm-Sub on Power's outputs to obtain the final result. 

\subsection{Utility Measurement Module} \label{sec:UtilityMeasurement}

\ldpkup adopts several utility metrics to calculate the errors between frequencies. We exemplify four utility metrics implemented in LDP$^3$: $\ell_1$ distance, $\ell_2$ distance, Kullback-Leibler Divergence, and Earth Mover's Distance (EMD). Note that although we present the utility metrics in a way that measures the difference between $f(v)$ and $\tilde{f}(v)$ below, it is also possible to utilize these metrics to measure the difference between $f(v)$ and $\hat{f}(v)$ or the difference between $\hat{f}(v)$ and $\tilde{f}(v)$.

\textbf{$\ell_1$ distance}, also known as Manhattan distance, measures error using the $\ell_1$ norm (absolute value norm). It is defined as:
\begin{equation}
    \text{$\ell_1$ distance} = \sum_{v \in \mathcal{D}} \left| \tilde{f}(v) - f(v)  \right|
\end{equation}

\textbf{$\ell_2$ distance}, also known as Euclidean distance, measures error using the $\ell_2$ norm (squared error norm). It is defined as:
\begin{equation}
    \text{$\ell_2$ distance} = \sqrt{ \sum_{v \in \mathcal{D}} \left( \tilde{f}(v) - f(v) \right)^2 }
\end{equation}

\textbf{Kullback-Leibler Divergence} (KL-divergence) is a measure of how one probability distribution diverges from a second. Here, $f(v)$ and $\tilde{f}(v)$ are treated as probability distributions:
\begin{equation}
    \text{KL-divergence} = \sum_{v \in \mathcal{D}} f(v) \log \left( \frac{f(v)}{\tilde{f}(v)} \right)
\end{equation}

\textbf{Earth Mover's Distance} (EMD) measures the minimum cost of transforming one distribution into another. It can be interpreted as the amount of work required to transform the original frequency distribution $f(v)$ into the post-processed distribution $\tilde{f}(v)$. Mathematically, it is defined as:
\begin{equation}
\text{EMD} = \min_{\pi} \sum_{v_1, v_2 \in \mathcal{D}} \pi(v_1, v_2) \cdot |v_1 - v_2|
\end{equation}
where $\pi(v_1, v_2)$ represents the amount of mass moved from $v_1$ to $v_2$, and $|v_1 - v_2|$ is the distance between $v_1$ and $v_2$.

\subsection{Multi-Threading Module}

In practice, finding which protocol and PP method yields the highest utility requires experimenting with many protocol and method combinations. Furthermore, due to the randomized nature of LDP, repeating each experiment multiple times is needed to obtain a reliable and statistically significant result. On the other hand, performing many experiments with many repetitions yields undesirably high execution times. 

To address this problem, we implemented LDP$^3$ in a multi-thread\-ed architecture as shown in Figure \ref{fig:multithread_system}. The process can be broken down into several key steps. First, the dataset is divided into $t$ equal-sized chunks, where $t$ is the number of threads. Each chunk is assigned to a separate processing thread. A unique random seed is generated for each thread to maintain independence between threads. Second, on each thread, individual user values in the chunk (denoted by $v_1$ to $v_n$ in Figure \ref{fig:multithread_system}) undergo LDP perturbation using the selected protocol (e.g., GRR, OLH, RAPPOR, etc.). Intra-thread results are aggregated and estimated within the thread's execution. Each thread executes in parallel and independently from the others. Third, \ldpkup collects estimation results from all threads and combines them via averaging. This results in the estimated frequencies denoted by $\hat{f}(v)$. Finally, $\hat{f}(v)$ undergo selected PP method(s) from the Post-Processing Module (Section \ref{sec:postprocess}). This results in the post-processed frequencies denoted by $\tilde{f}(v)$. The errors between $f(v)$ and $\tilde{f}(v)$ are calculated using metrics implemented in the Utility Measurement Module (Section \ref{sec:UtilityMeasurement}). 


We note two important remarks which are taken into consideration when designing the multi-threading module. First, by default, sizes of all $t$ chunks are equal. This ensures that the averaging performed by \ldpkup is suitable for obtaining $\hat{f}(v)$. If the sizes of the chunks are different, then weighted averaging (where each chunk is given a weight directly proportional to its size) would ensure correctness. Second, domain size $\mathcal{D}$, privacy budget $\varepsilon$, and other protocol parameters are treated as \textit{global} parameters which are fixed across all threads. This ensures that the protocol behavior remains consistent across all threads.

\subsection{Execution Module}

This module provides an interface for executing \ldpkup in practice. The general command structure to run \ldpkup is shown in Figure \ref{fig:command}. It can be observed that the command includes several options:
\begin{itemize}
    \item \texttt{-e EPSILON} is used to determine the privacy budget $\varepsilon$.
    \item \texttt{-p PROTOCOLS} specifies the LDP protocol(s) to use. Possible protocols are those implemented in the Protocol Module of \ldpkup (Section \ref{sec:LDPprotocols}). One or more protocols can be used in the same execution. Specifying \texttt{"all"} for this option prompts \ldpkup to repeat the experiments with all available protocols in the Protocol Module.
    \item \texttt{-m METHODS} specifies the PP method(s) to use. Possible methods are those implemented in the Post-Processing Module (Section \ref{sec:postprocess}). One or more PP methods can be used in the same execution. Specifying \texttt{"all"} for this option prompts \ldpkup to repeat the experiments with all available methods.
    \item \texttt{-r REPEAT}: \ldpkup performs each experiment multiple times (i.e., multiple repetitions) to tackle the inherent randomness of LDP and to achieve statistical significance. The \texttt{-r} option is used to determine the number of repetitions per experiment, e.g., 10.
    \item \texttt{-t THREAD\_NUMBER} specifies the number of parallel threads to use in the multi-threaded execution of LDP$^3$.
    \item \texttt{-d DATASET} specifies the dataset path, e.g., \texttt{my\_dataset.csv}. Currently, it is expected that datasets will contain rows of values where each row corresponds to one user's data. The code in \ldpkup which reads and parses the datasets can be modified to support datasets with different formats.
    \item \texttt{-u UTILITY\_METRIC} specifies the utility metric to evaluate errors, selected among those implemented in the Utility Measurement Module (Section \ref{sec:UtilityMeasurement}). 
\end{itemize}

\section{Experimental Evaluation} \label{sec:Experiments}

\subsection{Experiment Setup}

In this section, we perform experiments to demonstrate that: (i) to improve utility, it is highly beneficial to run experiments with different LDP protocol and PP method combinations using \ldpkup to find a good combination, rather than using a fixed protocol or PP method; and (ii) multi-threading in \ldpkup brings substantial benefits that speed up experiment execution. 

\begin{figure}[!t]
\centering
\vspace{6pt}
\begin{lstlisting}[language=bash]
python3 main.py -e EPSILON -p PROTOCOLS  -m METHODS -r REPEAT -t THREAD_NUMBER -d DATASET -u UTILITY_METRIC
\end{lstlisting}
\vspace{-2pt}
\caption{Command structure to run \ldpkup with various options and parameters.}
\vspace{-4pt}
\label{fig:command}
\end{figure}

We used three real-world datasets for experimentation: BMS-POS, Kosarak, and Porto. We obtained Kosarak from the SPMF Dataset Repository\footnote{\url{www.philippe-fournier-viger.com/spmf/index.php?link=datasets.php}}, BMS-POS from the public Github repository\footnote{\url{https://github.com/cpearce/HARM/blob/master/datasets/BMS-POS.csv}}, and Porto is from the ECML-PKDD Taxi Service Prediction Challenge. 
\begin{itemize}
    \item Kosarak contains click-stream data of a Hungarian online news portal. Due to many URLs having very few occurrences (e.g., one or two), we pre-processed the dataset by identifying the top-128 most visited URLs and removed the rest, i.e., $|\mathcal{D}| = 128$. For users who had more than one URL in their resulting stream, the most frequently occurring URL in their stream was picked as their $v_u$.
    \item BMS-POS includes market basket sales data from a major electronics retailer, consisting of 515,596 transactions and 1,657 unique items sold. In our experiments, we pre-processed the dataset in a similar fashion to Kosarak by keeping only the top 256 most frequently purchased items.
    \item Porto contains trips of 442 taxis driving in the city of Porto. While the dataset contains full taxi trips, we pre-processed it by keeping only the starting location of each trip and applied a $15 \times 15$ grid for discretization. Each trip was treated as a new $v_u$. Consequently, we have $|\mathcal{D}|$ = 225 and $|\mathcal{U}|$ = 1,620,157.
\end{itemize}
Each experiment was conducted 10 times on an Intel Alder Lake Core i7 1255U CPU and the average results are reported. We use $\ell_1$ distance as the utility metric. 


\begin{table*}[!h]
    \centering
    \caption{$\ell_1$ distances of estimations under different combinations of LDP protocols and post-processing methods. Kosarak dataset and $\varepsilon$ = 1 are used. All values in the table are $\times 10^{-3}$.}
    \label{tab:protocol_comparison_kosarak}
    \scriptsize 
    \resizebox{\textwidth}{!}{ 
    \begin{tabular}{|c|c|c|c|c|c|c|c|c|c|} 
\hline
\textbf{Protocol} & \textbf{w/o PP} & \textbf{Avg. w/ PP} & \textbf{Base-Pos} & \textbf{Norm} & \textbf{Norm-Cut} & \textbf{Norm-Mul} & \textbf{Norm-Sub} & \textbf{Power} & \textbf{PowerNS}  \\ 
\hline
GRR               & 5.65            & 3.83                & 3.87              & 5.65          & 3.53              & 3.98              & 3.08              & 3.66           & \underline{3.02}              \\
OLH               & 1.56            & 1.44                & 1.33              & 1.56          & 1.47              & 1.39              &  \underline{1.27}              & 1.59           & 1.49              \\
BLH               & 1.84            & 1.65                & 1.48              & 1.83          & 1.71              & 1.58              &  \underline{1.39}              & 1.84           & 1.70              \\
OUE               & 1.48            & 1.35                & 1.23              & 1.47          & 1.35              & 1.27              &  \underline{1.19}              & 1.50           & 1.46              \\
RAPPOR            & 1.66            & 1.51                & 1.37              & 1.65          & 1.52              & 1.47              &  \underline{1.30}              & 1.71           & 1.61              \\
SS                & 1.60            & 1.39                & 1.34              & 1.60          & 1.49              & 1.43              & 1.28              & 1.32           & \underline{1.21}        \\
\hline
\end{tabular}
    }
\end{table*}

\begin{table*}[!h]
    \centering
    \caption{$\ell_1$ distances of estimations under different combinations of LDP protocols and post-processing methods. BMS-POS dataset and $\varepsilon$ = 1 are used. All values in the table are $\times 10^{-3}$.}
    \label{tab:protocol_comparison_BMS_POS}
    \scriptsize 
    \resizebox{\textwidth}{!}{ 
\begin{tabular}{|c|c|c|c|c|c|c|c|c|c|} 
\hline
\textbf{Protocol} & \textbf{w/o PP} & \textbf{Avg. w/ PP} & \textbf{Base-Pos} & \textbf{Norm} & \textbf{Norm-Cut} & \textbf{Norm-Mul} & \textbf{Norm-Sub} & \textbf{Power} & \textbf{PowerNS}  \\ 
\hline
GRR               & 10.79           & 5.94                & 6.74              & 10.79         & 5.15              & \underline{4.01}              & 4.32              & 6.36           & 4.20              \\
OLH               & 2.14            & 2.00                & 1.76              & 2.14          & 1.98              & 1.67              &  \underline{1.66}              & 2.52           & 2.26              \\
BLH               & 2.45            & 2.24                & 1.95              & 2.45          & 2.19              &  \underline{1.83}              & 1.84              & 2.91           & 2.53              \\
OUE               & 2.17            & 2.03                & 1.78              & 2.17          & 1.98              &  \underline{1.71}   &  \underline{1.71}              & 2.53           & 2.30              \\
RAPPOR            & 2.21            & 2.02                & 1.81              & 2.21          & 2.01              &  \underline{1.73}   &  \underline{1.73}              & 2.44           & 2.23              \\
SS                & 2.08            & 1.79                & 1.70              & 2.08          & 1.92              & 1.62              & 1.61              & 1.74           &   \underline{1.57}              \\
\hline
\end{tabular}
    }
\end{table*}

\begin{table*}[!h]
    \centering
    \caption{$\ell_1$ distances of estimations under different combinations of LDP protocols and post-processing methods. Porto dataset and $\varepsilon$ = 1 are used. All values in the table are $\times 10^{-3}$.}
    \label{tab:protocol_comparison_porto}
    \scriptsize 
    \resizebox{\textwidth}{!}{ 
\begin{tabular}{|c|c|c|c|c|c|c|c|c|c|} 
\hline
\textbf{Protocol} & \textbf{w/o PP} & \textbf{Avg. w/ PP} & \textbf{Base-Pos} & \textbf{Norm} & \textbf{Norm-Cut} & \textbf{Norm-Mul} & \textbf{Norm-Sub} & \textbf{Power} & \textbf{PowerNS}  \\ 
\hline
GRR               & 5.66           & 3.97                & 3.84              & 5.66         & 3.45              & 3.19              & \underline{3.03}              & 4.33           & 4.33              \\
OLH               & 1.17            & 1.06                & 0.93              & 1.17          & 0.93              & 0.94              &  \underline{0.87}              & 1.27           & 1.27              \\
BLH               & 1.34            & 1.19                & 1.05              & 1.34          & 1.04              &  1.06              & \underline{0.97}              & 1.45           & 1.45              \\
OUE               & 1.21            & 1.09                & 0.96              & 1.21          & 0.97              &  0.97   &  \underline{0.90}              & 1.33           & 1.33              \\
RAPPOR            & 1.27            & 1.14                & 1.01              & 1.27          & 1.01              &  1.01   &  \underline{0.94}              & 1.39           & 1.39              \\
SS                & 1.18            & 0.94                & 0.94              & 1.18          & 0.94              & 0.95              & 0.88              & \underline{0.86}           &   \underline{0.86}              \\
\hline
\end{tabular}
    }
\end{table*}

\subsection{Utility Benefits of \ldpkup}

In the first set of experiments, we demonstrate the utility benefits of \ldpkup by reporting the errors in frequency estimations without post-processing (w/o PP), with each post-processing method from Section \ref{sec:postprocess}, and the average of all post-processing methods. Results with the Kosarak dataset are shown in Table \ref{tab:protocol_comparison_kosarak}, BMS-POS dataset are shown in Table \ref{tab:protocol_comparison_BMS_POS}, and Porto dataset are shown in Table \ref{tab:protocol_comparison_porto}. For each protocol (each row), the lowest error among all post-processing methods is underlined. 

We observe that in many cases, errors are high when there is no post-processing. Post-processing methods consistently help in reducing errors. This consistent improvement demonstrates the effectiveness of post-processing methods and motivates the utility of \ldpkup in enabling the convenient integration and execution of various post-processing methods. Additionally, different post-processing methods yield different amounts of improvement, and the optimal post-processing method varies depending on the protocol and dataset. In other words, there is no post-processing method that performs universally best across all protocols and datasets. For example, in Table \ref{tab:protocol_comparison_BMS_POS}, Norm-Mul is best for GRR, Norm-Sub is best for OLH, and PowerNS is best for SS. Furthermore, choosing the best protocol and post-processing method is indeed important since the best choice (e.g., SS - PowerNS combination) yields substantially lower error (1.57) compared to a bad choice or an average choice (e.g., $>$ 2.0 error). By offering a range of post-processing methods, \ldpkup can empower researchers and practitioners to experiment with multiple configurations, enabling them to find the best combination of LDP protocol and post-processing method for their specific requirements. 


\subsection{Benefits of Multi-Threading} 

\begin{figure}[!t]
\centering
    \includegraphics[width=.85\columnwidth]{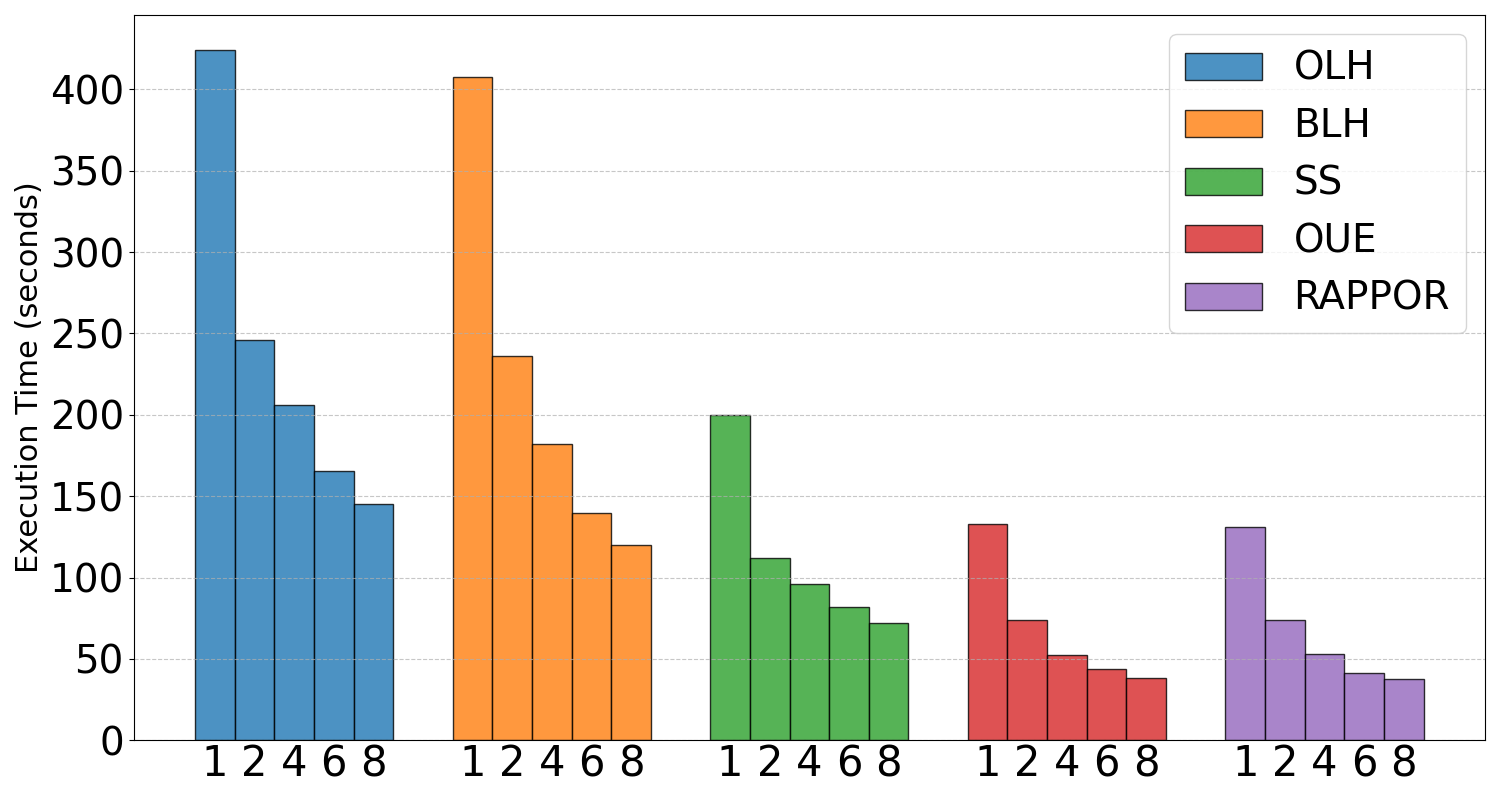}
    \caption{Execution time vs number of threads (Kosarak).}
    \label{fig:kosarak_multithread}
\end{figure}

\begin{figure}[!t]
\centering
    \includegraphics[width=.85\columnwidth]{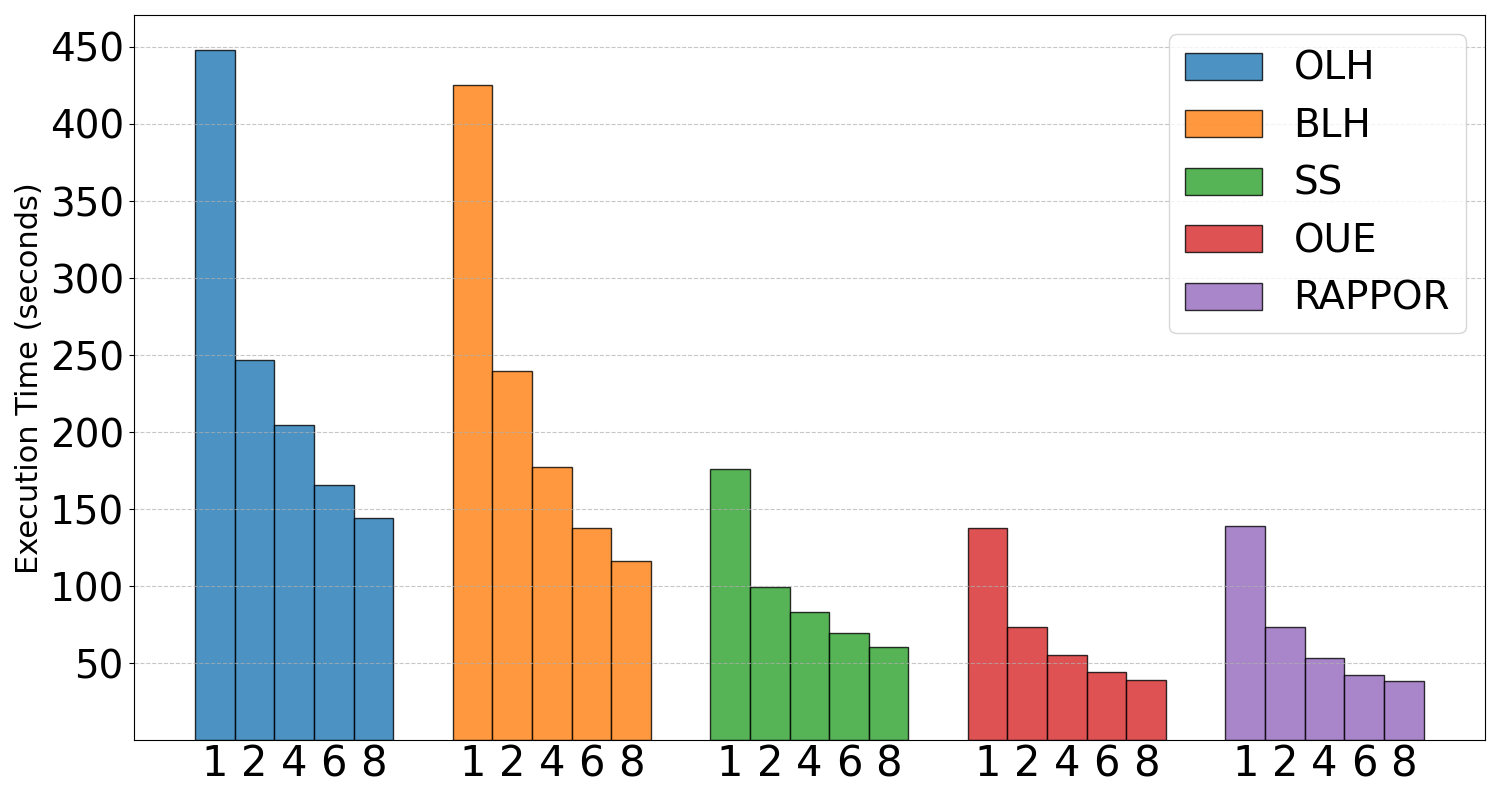}
    \caption{Execution time vs number of threads (BMS-POS).}
    \label{fig:bms_multithread}
\end{figure}

\begin{figure}[!t]
\centering
    \includegraphics[width=.85\columnwidth]{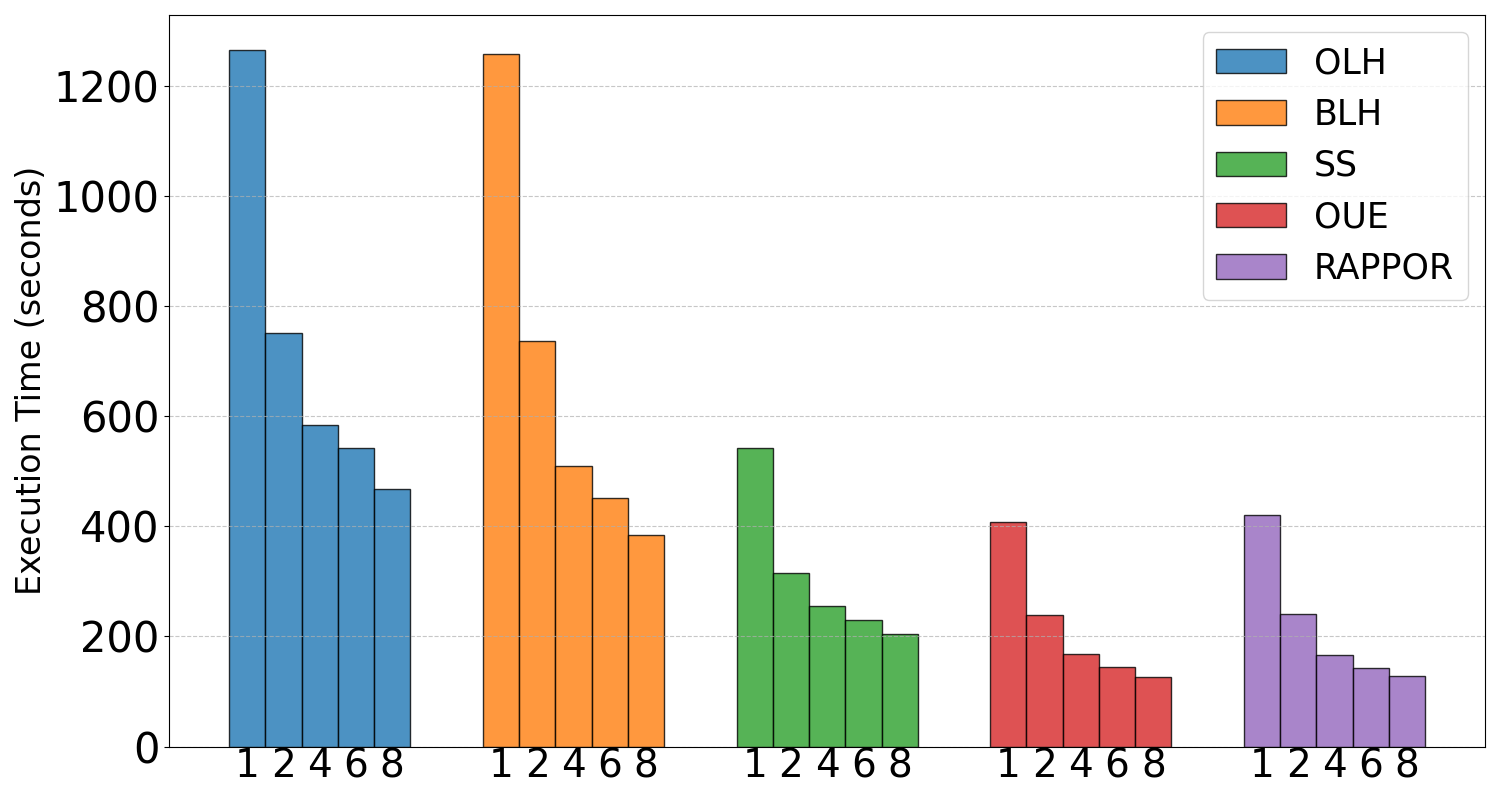}
    \caption{Execution time vs number of threads (Porto).}
    \label{fig:porto_multithread}
\end{figure}

In the second set of experiments, we demonstrate the benefits of the multi-threaded design of \ldpkup by evaluating the impact of multi-threading on the execution times of various protocols. Figures \ref{fig:kosarak_multithread}, \ref{fig:bms_multithread}, and \ref{fig:porto_multithread} show the total time required to run each experiment (10 repetitions) on the Kosarak, BMS-POS, and Porto datasets, respectively. 

As expected, we observe that the execution times decrease as we increase the number of threads from 1 to 8. However, the decrease is not linear. This is because, as shown in Figure \ref{fig:multithread_system}, \ldpkup needs to divide the input dataset into $t$ chunks at the beginning. In addition, it needs to combine the estimation results from all threads (chunks) and apply post-processing to the combined estimation result. These actions cannot be multi-threaded; therefore, their time cost cannot be eliminated or reduced by increasing the number of threads. On the other hand, for many protocols, the total execution times are halved or reduced to one-third as we go from a single thread to 4 or more threads. The benefits of multi-threading are especially noticeable in cases where the protocol's execution times are generally high (such as OLH, BLH). Overall, these results highlight the benefits of LDP$^3$'s multi-threading and its suitability for large-scale experimentation. 


\section{Conclusion} \label{sec:Conclusion}

In this paper, we introduced LDP$^3$, an extensible, open-source, and modular toolkit designed to advance research and practical applications of local differential privacy (LDP). By integrating widely used LDP protocols, post-processing methods, and utility metrics within a multi-threaded framework, \ldpkup provides researchers and practitioners with a powerful resource for benchmarking and testing existing or newly proposed methods, as well as exploring the optimized combinations protocols and post-processing methods suitable for their task. Our experimental results highlight the significant utility and efficiency gains enabled by LDP$^3$. We hope that \ldpkup will be helpful in accelerating LDP research and deployment, as well as fostering collaboration and reproducibility through the public open-source repository. 


\section*{Acknowledgment}

This study was supported by Scientific and Technological Research Council of Türkiye (TUBITAK) under Grant Number 123E179. The authors thank TUBITAK for their support.

\bibliographystyle{IEEEtran}
\bibliography{references}

\end{document}